\documentclass[usenatbib]{article}
\usepackage{times,graphicx,natbib,a4wide, hyperref}

\begin{document}
	
\title{Predator-Prey Behaviour in Self-Replicating Interstellar Probes}
\author{Duncan H. Forgan$^1$}
\maketitle

\noindent $^{1}$Centre for Exoplanet Science, SUPA, School of Physics \& Astronomy, University of St Andrews, St Andrews KY16 9SS, UK \\

\noindent \textbf{Word Count: 3,693} \\

\noindent \textbf{Direct Correspondence to:} \\
D.H. Forgan \\ \\
\textbf{Email:} dhf3@st-andrews.ac.uk \\
\textbf{Post:} Dr Duncan Forgan \\
Astronomy Suite, School of Physics and Astronomy
University of St Andrews \\
North Haugh, St Andrews \\
KY16 9SS, UK

\newpage

\begin{abstract}

\noindent The concept of a rapid spread of self-replicating interstellar probes (SRPs) throughout the Milky Way adds considerable strength to Fermi's Paradox.  A single civilisation creating a single SRP is sufficient for a fleet of SRPs to grow and explore the entire Galaxy on timescales much shorter than the age of the Earth - so why do we see no signs of such probes? One solution to this Paradox suggests that self-replicating probes eventually undergo replication errors and evolve into predator-prey populations, reducing the total number of probes and removing them from our view.  

I apply Lotka-Volterra models of predator-prey competition to interstellar probes navigating a network of stars in the Galactic Habitable Zone to investigate this scenario.  I find that depending on the local growth mode of both populations and the flow of predators/prey between stars, there are many stable solutions with relatively large numbers of prey probes inhabiting the Milky Way.  The solutions can exhibit the classic oscillatory pattern of Lotka-Volterra systems, but this depends sensitively on the input parameters.  Typically, local and global equilibria are established, with prey sometimes outnumbering the predators.  Accordingly, we find this solution to Fermi's Paradox does not reduce the probe population sufficiently to be viable.


\end{abstract}

\section{Introduction}

\noindent Why have we detected no sign of intelligent life beyond the Earth? This fundamental question continues to challenge our deepest-held beliefs about humanity and our place in the Universe.  Fermi's Paradox forces us to confront our Copernican assumptions about our lack of uniqueness with the lack of extraterrestrial intelligences \citep[ETIs, see e.g.][]{BrinG.D.1983, Cirkovic2009}.  Its strongest formulation can be given as follows \citep{Tipler1980}.  

Imagine a civilisation constructs an interstellar probe that is self-replicating.  Such a probe would be able to produce a copy every time it visits a new star system.  As each copy makes copies, the number of self-replicating probes (SRPs) grows exponentially, and every star in the Milky Way is explored on a timescale much, much shorter than its age.  Estimates for this exploration timescale vary, but are as short as ten million years \citep{Nicholson2013}, and perhaps shorter still. 

Given that this timescale is much shorter than the age of the Earth, and only one ETI constructing SRPs is sufficient to produce this scenario, on balance we should expect to see an interstellar probe orbiting the Sun.  And yet, we do not.  How can this be resolved? 

Among many possibilities, we can include solutions that require civilisations to be rare.  However, as a single civilisation is sufficient to swamp the galaxy in SRPs, we are effectively asking for humanity to be alone in the Universe.

It may well be the case that other intelligent beings exist, and that their probes are \emph{en route}, and may not arrive for several thousand or several million years.  This demands that the biological timescale of most of the Milky Way is somehow correlated.  Perhaps this is due to global regulation mechanisms, large scale destructive events that reset the biological clocks of many civilisations simultaneously \citep{Annis,Vukotic_and_Cirkovic_07,Vukotic_and_Cirkovic_08}.  However, there are no known astrophysical regulation mechanisms that are truly global.  For example, when the Milky Way's central supermassive black hole enters an active accreting phase, the subsequent radiation output results in a regulation mechanism with a correlation length of order ten thousand light years \citep{Balbi2017}.  The Milky Way is much larger than this correlation length, and even a small handful of uncorrelated biospheres is fit to make this type of solution untenable.

Of course, it may well be the case that an interstellar probe \emph{is} in the outer solar system, and we have not yet found it \citep{Papagiannis1978,Freitas1983}.  While this is indeed plausible, the possible places for an interstellar probe to hide continues to decrease \citep[see also][]{Haqq-Misra2012}.

Other solutions suggest that SRPs themselves are rare.  There have been several arguments put forth regarding the safety of SRPs as a technology.  \citet{Sagan1983} suggest that civilisations will voluntarily refrain from building SRPs, for a host of reasons.  For example, self-replication could result in encoding errors.  These ``mutations'' will propagate from generation to generation with unforeseen, unintended consequences.  The consequences of this could be severe, e.g. a genocidal conversion of a species and its technology into probes.  It has been argued that these risks would persuade intelligent beings to place a moratorium on SRPs.  Such arguments are notorious for their anthropological assumptions, and the larger assumption that ``civilisations'' are unified in purpose and execution - an assumption humanity repeatedly invalidates (for a limited set of examples, see e.g. \citealt{Collins2008, Denning2011,Lempert2014}, also \citealt{Forgan2017d}).  This heterogeneity crucially undermines the likelihood of a Galactic moratorium.  Once technology achieves a certain threshold, a very small number of individuals in a single civilisation become capable of building illegal technology and producing a species-ending moment, or indeed an SRP fleet \citep[see e.g.][]{Sotos2017}.

We will explore a variant of the unintended consequences of SRPs, the ``Predator-Prey'' hypothesis \citep{Chyba2005}.  In this scenario, a subset of SRPs mutate into predators of other SRPs.  For an SRP to make a copy of itself, it is likely that cannibalising another SRP will be the most energy-efficient solution.  In the standard description of this scenario, predators reduce the prey population until the available prey are exhausted, reducing the visibility of SRPs in the Milky Way.

However, we should also expect in this scenario that the spread of SRPs across interstellar space will be modulated by the non-trivial population dynamics of predator-prey systems, which are among the most well-studied fundaments of mathematical biology.  The full implications of these population dynamics in SRPs are largely unstudied (although see \citealt{Wiley2011}).

In this work, we apply the classic Lotka-Volterra formalism for predator-prey systems to an interstellar network.  In the network, each star plays host to a predator-prey population, which transmits and receives both predator and prey from neighbouring star systems.  We will consider under what circumstances predators can reduce the population significantly, as well as what circumstances permit stable, significant populations of prey to remain in existence.


\section{Method}

\begin{table*}
\centering
\caption{A list of variables used in this paper, with their definition. \label{tab:var}}
\begin{tabular}{c|c}
\hline
\hline
$R_i$ & Number of prey at star $i$  (thousands)\\
$b_{R,i}$ & Birth rate of prey at star $i$ (thousands/Myr) \\
$d_{R,i}$ & Death rate of prey at star $i$ (thousands/Myr) \\
$K_{R,i}$ & Carrying capacity of prey at star $i$ (thousands) \\
$I_{R,i}$ & Inflow of prey to star $i$ (thousands/Myr) \\
$O_{R,i}$ & Outflow of prey from star $i$ (thousands/Myr)\\
$o_{R,i}$ & Outflow rate of prey from star $i$ (thousands/Myr)\\
\hline
$P_i$ & Number of predators at star $i$ (thousands) \\
$b_{P,i}$ & Birth rate of predators at star $i$ (thousands/Myr) \\
$d_{P,i}$ & Death rate of predators at star $i$ (thousands/Myr) \\
$K_{R,i}$ & Carrying capacity of prey at star $i$ (thousands) \\
$I_{P,i}$ & Inflow of predators to star $i$ (thousands/Myr)\\
$O_{P,i}$ & Outflow of predators from star $i$ (thousands/Myr)\\
 \hline
 \hline
\end{tabular}
\end{table*}

\noindent We model the Galactic Habitable Zone (GHZ) as a graph of $N_*$ stars.  The stars are distributed in space, with their locations fixed.  The semimajor axes of the stars $a_i$ around the Galactic Centre are exponentially distributed to simulate the Milky Way's surface density profile:

\begin{equation} 
P(a_i) \propto e^{-\frac{a_i}{r_S}},
\end{equation}

\noindent with the scale radius $r_S=3.5$ kpc \citep{Ostlie_and_Caroll}. The minimum and maximum permitted radius of the stars is $[7,10]$ kpc respectively, following the \citet{Gowanlock2011} model of the GHZ.  We assume a uniform eccentricity distribution, provided that the star's orbit prevents its closest approach to the Galactic Centre being smaller than the inner radius of the GHZ.  We also restrict the inclination of the orbits so that they do not exceed 0.5 radians.  The longitude of the ascending node, the argument of periapsis and the true anomaly are uniformly sampled in the range $[0,2\pi]$ radians.  Note that once the stellar orbital parameters are determined, we fix the stellar positions throughout the calculation.

\begin{figure}
\begin{center}
\includegraphics[width=0.7\textwidth]{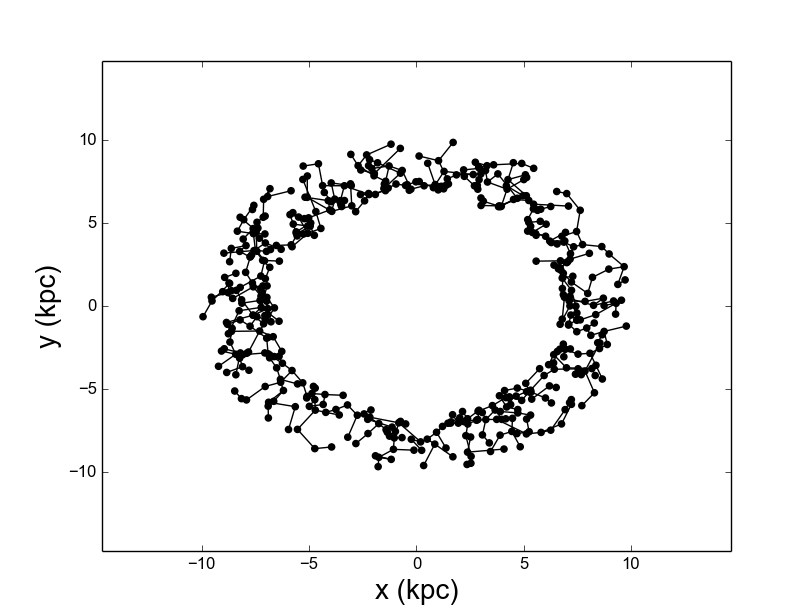}
\end{center}
\caption{The stellar network on which we conduct our calculations.  Each star represents an individual system on which we integrate the Lotka-Volterra equations.  Lines indicate edge connections between stars, which determines the rate of outflow/inflow of predators and prey onto each star (see text).  This stellar network is the minimum spanning tree $T$ of the graph $G$ (see text for definition).} 
\label{fig:network}
\end{figure}

Each star provides a vertex to our graph, and we construct the graph $G$ such that for a star $i$, any star $j$ within a minimum distance $R_{\rm min}$ of $i$ is connected by an edge.  We then define the minimum spanning tree $T$ of this graph, and use this for computing predator/prey evolution (see Figure \ref{fig:network}).

For each star $i$ in the network, we solve the Lotka-Volterra equations for the local numbers of prey $R_i$, and number of predators $P_i$, with inflow and outflow rates of prey determined by the edges that connect each star to its neighbours.   For logistic growth, the equations for star $i$ are:

\begin{eqnarray}
\frac{dR_i}{dt} = b_{R,i} R_i \left( 1- \frac{R_i}{K_{R,i}}\right) -d_{R,i} R_i P_i - O_{R,i} + I_{R,i} \\
\frac{dP_i}{dt} = b_{P,i} P_i R_i  -d_{P,i} P_i - O_{P,i} + I_{P,i} \\
\end{eqnarray}

\noindent To model exponential growth, we can simply set the carrying capacity $K$ to very large values ($10^{30}$).  We give a list of all variables with their definition in Table \ref{tab:var}.  The outflow of prey from star $i$ to star $j$, $O_{R,ij}$ is calculated assuming a fixed outflow rate $o_{R,i}$, a fixed probe speed $v_{\rm probe}$, and the distance between $i$ and $j$, $D_{ij}$:

\begin{equation}
O_{R,ij} = \frac{o_{R,i} v_{\rm probe}}{D_{ij}} R_i 
\end{equation}

\noindent As we begin our simulations with many stars containing zero prey initially, we demand that $O_{R,ij}$ be zero until a sufficient time has elapsed for the first prey to arrive at $i$ to complete the journey to $j$.  The total outflow is then

\begin{equation}
O_{R,i} = \sum_j O_{R,ij}
\end{equation}

\noindent Inflows are computed similarly:

\begin{equation}
I_{R,i} = \sum_j I_{R,ij}
\end{equation}

\noindent where all outflow/inflow terms are pairwise, i.e.

\begin{equation}
I_{R,ji} = O_{R,ij} 
\end{equation}

\noindent and an identical set of equations are used for predator outflow and inflow $\left(O_{P,i},I_{P,i}\right)$.

\section{Results}

\subsection{Tests}

\subsubsection{Single Star, Exponential Growth \label{sec:exp_singlestar}}

\noindent We test the code by considering a single star with no inflow or outflow, to ensure that we retrieve the solution to the classic or ``vanilla'' Lotka-Volterra equations for exponential growth (i.e. infinite carrying capacity), where we now drop the $i$ subscript for clarity:

\begin{eqnarray}
\frac{dR}{dt} = b_{R} R - d_{R} R P \\
\frac{dP}{dt} = b_{P} P R - d_{P} P \\
\end{eqnarray}

\begin{table*}
\centering
\caption{Parameter values used for the results shown in Figure \ref{fig:exp_singlestar}. \label{tab:exp_singlestar}}
\begin{tabular}{c|c}
\hline
\hline
$R$ (initial) & 1.8 \\
$b_{R}$ & 0.6\\
$d_{R}$ & 1.333 \\
\hline
$P$ (initial) & 1.0 \\
$b_{P}$ & 1.0 \\
$d_{P}$ & 1.0 \\
 \hline
 \hline
\end{tabular}
\end{table*}

\begin{figure*}
\begin{center}
\includegraphics[width=0.49\textwidth]{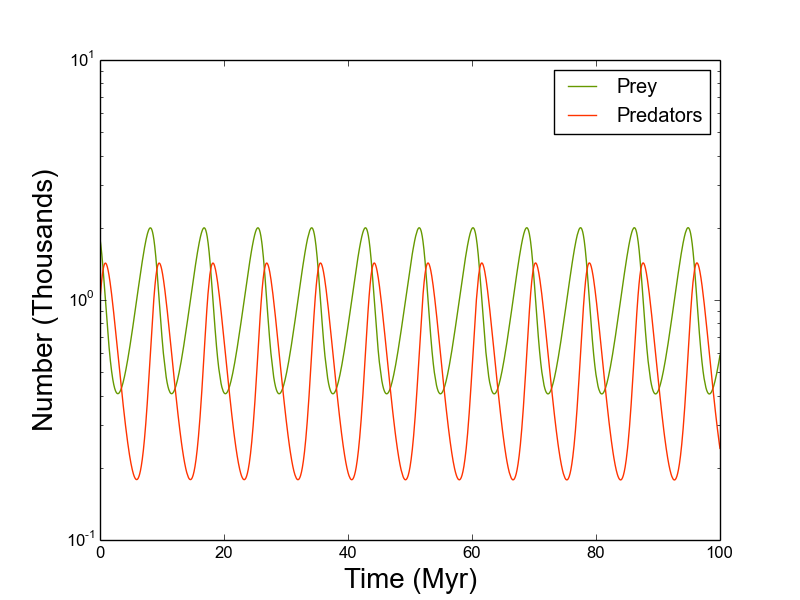}
\includegraphics[width=0.49\textwidth]{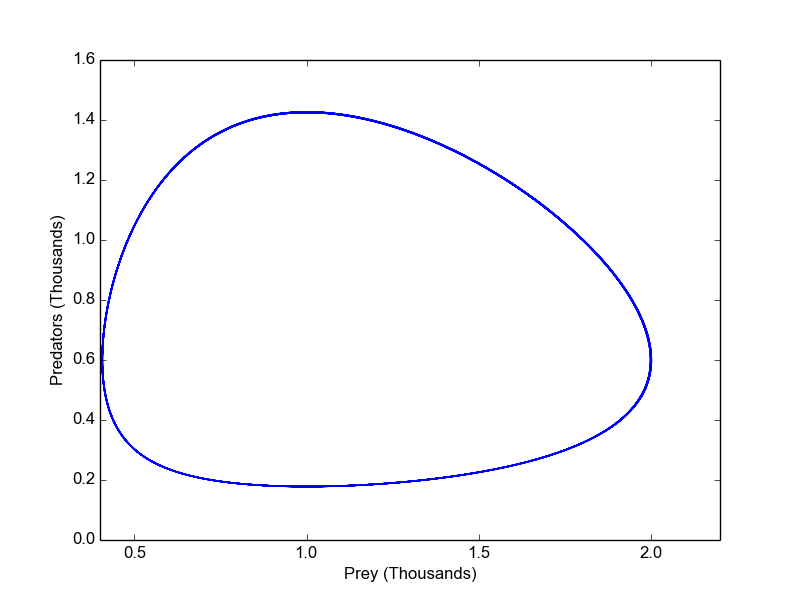}
\end{center}
\caption{The Lotka-Volterra solution assuming exponential growth in a single star system with zero inflow/outflow.}
\label{fig:exp_singlestar}
\end{figure*}

\noindent Table \ref{tab:exp_singlestar} shows the parameter values for this run, and Figure \ref{fig:exp_singlestar} shows the resulting behaviour.  We see that the predator/prey populations are oscillatory, both with period 8.6 Myr, and out of phase. This represents a fixed locus in predator-prey space (right panel of Figure \ref{fig:exp_singlestar}).  This locus can be determined by condensing the coupled equations into a single equation:

\begin{equation}
\frac{d}{dt} \left(b_{P} R + d_{R} P -  d_{P} \log R - b_{R} \log P\right) = 0 
\end{equation}

\noindent If we define the above as a Hamiltonian:

\begin{equation}
H(R,P) = b_{P} R_i + d_{R} P -  d_{P} \log R - b_{R} \log P 
\end{equation}

We can use standard Hamiltonian analysis (see \citealt{Murray2004}) to obtain a single solution for the predator/prey population:

\begin{equation}
(R,P) = \left(\frac{d_{P}}{b_{P}}, \frac{b_{R}}{d_{R}}\right).
\end{equation}

\noindent This defines an initial point on the $R-P$ locus, with all other points on the locus defining a surface of constant $H$.

\subsubsection{Single Star, Logistic Growth \label{sec:log_singlestar}}

\begin{table*}
\centering
\caption{Parameter values used for the results shown in Figure \ref{fig:log_singlestar}. \label{tab:log_singlestar}}
\begin{tabular}{c|c}
\hline
\hline
$R$ (initial) & 1.8 \\
$b_{R}$ & 0.6\\
$d_{R}$ & 1.333 \\
$K_{R}$ & 20 \\
\hline
$P$ (initial) & 1.0 \\
$b_{P}$ & 1.0 \\
$d_{P}$ & 1.0 \\
 \hline
 \hline
\end{tabular}
\end{table*}

\begin{figure*}
\begin{center}
\includegraphics[width=0.49\textwidth]{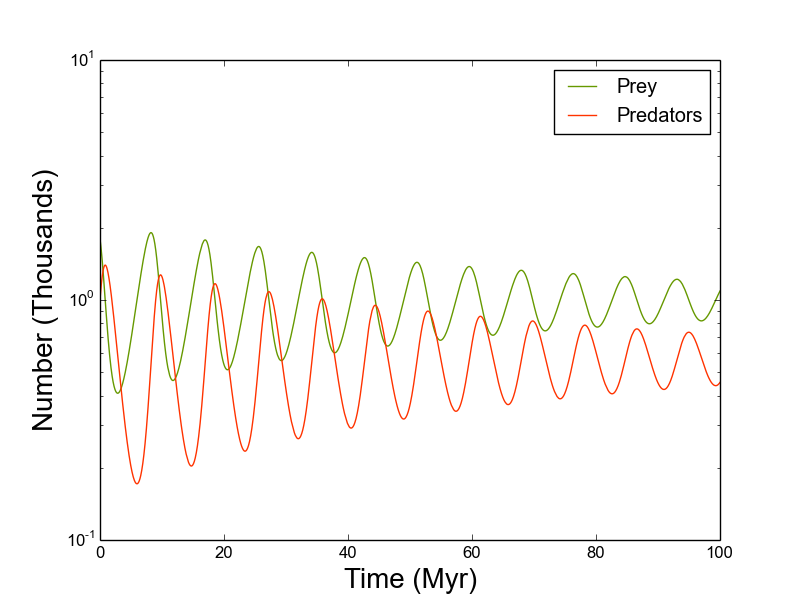}
\includegraphics[width=0.49\textwidth]{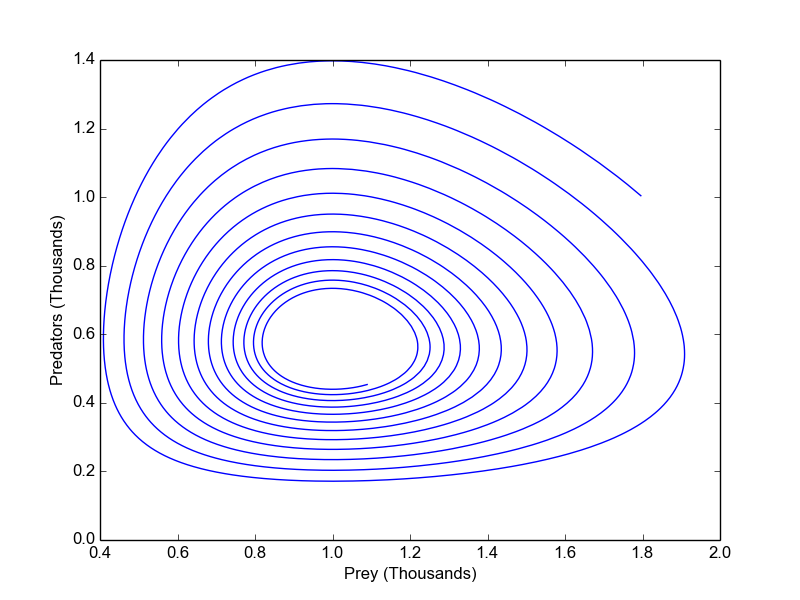}
\end{center}
\caption{The Lotka-Volterra solution assuming logistic growth in a single star system with zero inflow/outflow.  The prey carrying capacity is 20.}
\label{fig:log_singlestar}
\end{figure*}

\noindent We repeat the calculation of section \ref{sec:exp_singlestar}, where we now impose a prey carrying capacity $K_{R}=20$ (see Table \ref{tab:log_singlestar}).  We recover the standard result, that the stabilising effect of carrying capacity damps the oscillations in predator/prey populations, until an equilibrium is eventually found at late times (Figure \ref{fig:log_singlestar}). This is represented in $R-P$ space by a spiral with end point given by the equilibrium solution.  Note that the equilibrium prey value is much less than the carrying capacity, which is a common outcome in models of this type.  Depending on the input parameters, steady state solutions are possible where either the prey or predators dominate the combined population.

\subsection{Logistic Growth, Globally Constant Parameters }

\subsubsection{Moderate, Constant Outflow Rate \label{sec:log_constant}}

\begin{table*}
\centering
\caption{Parameter values used for the results shown in section \ref{sec:log_constant}. \label{tab:log_constant}}
\begin{tabular}{c|c}
\hline
\hline
$R_1$ (initial) & 1.8 \\
$b_{R,i}$ & 0.6\\
$d_{R,i}$ & 1.33 \\
$K_{R,i}$ & 20 \\
$o_{R,i}$ & $10^{-3}$ \\
\hline
$P_1$ (initial) & 1.0 \\
$b_{P,i}$ & 1.0 \\
$d_{P,i}$ & 1.0 \\
 \hline
 \hline
\end{tabular}
\end{table*}

\begin{figure}
\begin{center}
\includegraphics[width=0.7\textwidth]{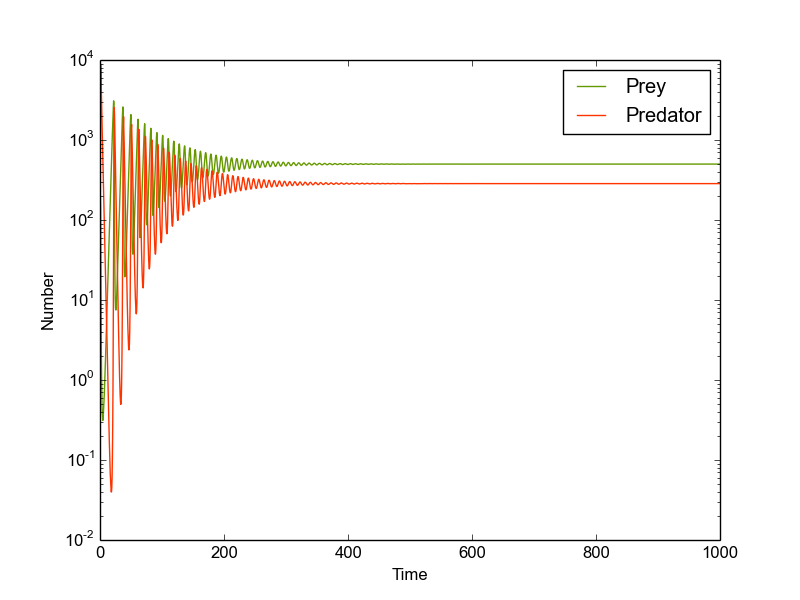}
\end{center}
\caption{The predator/prey population in the Galactic Habitable Zone, assuming logistic growth over 500 stars with globally fixed predator/prey growth and death rates (see Table \ref{tab:log_constant}).}
\label{fig:log_constant}
\end{figure}

\noindent We now consider a full stellar network, where each star possesses the same fixed values for all parameters (see Table \ref{tab:log_constant}).  We place an initial population $(R_{1},P_{1})$ on star 1, with all other stars hosting zero prey/predators initially.  The global picture is similar to the single star case (section \ref{sec:log_singlestar}) - the initial oscillatory phase is quickly damped towards equilibrium values.  Again, we find that the total prey population is lower than the maximum permitted by carrying capacity ($N_* K_{R,i}=10^4$), but the predator population is also constrained, and hence we find the prey population can be sustained at relatively high levels (provided they do not exhaust local resources for self-replication).

Interrogating individual stars reveals that all systems assume damped oscillatory states similar to the previous section, all with equal oscillation periods for both the prey and predator populations (now 9.17 Myr).  This is made possible by the relatively large outflow rate ($o_{R,i}=10^{-3}$), which seeds a system with sensible initial quantities of prey/predators, while remaining sufficiently small that the internal predator-prey dynamics dominates the population's evolution.

As the initial prey population requires time to traverse the stellar network, each predator-prey system begins operating at a slightly different initial time.  The oscillations seen in the global population have a slightly increased wavelength compared to individual systems, due to the constructive interference of many oscillatory curves, each slightly out of phase with each other, ``smearing'' the curve over a longer period range.

\subsubsection{Low, Constant Outflow Rate}

\begin{figure}
\begin{center}
\includegraphics[width=0.7\textwidth]{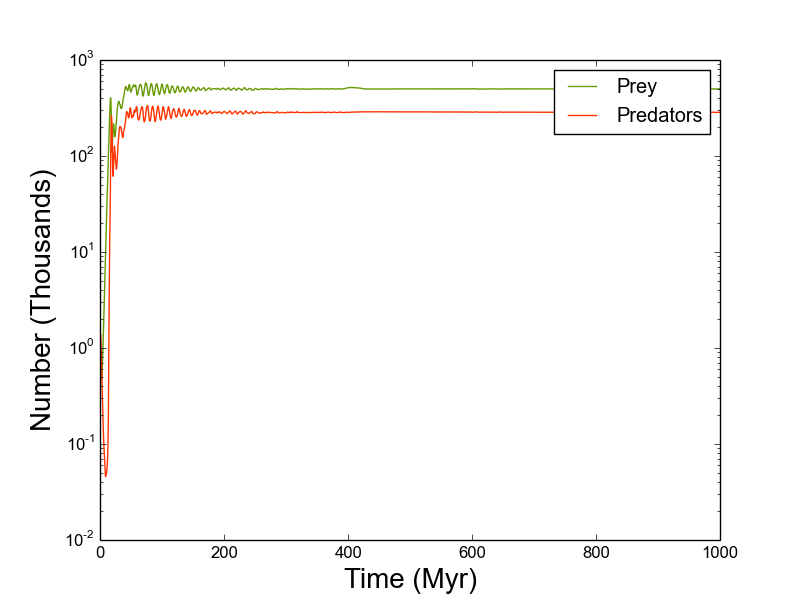}
\end{center}
\caption{As Figure \ref{fig:log_constant}, but with significantly reduced outflow.}
\label{fig:log_verylowoutflow}
\end{figure}

\begin{figure*}
\begin{center}
\includegraphics[width=0.49\textwidth]{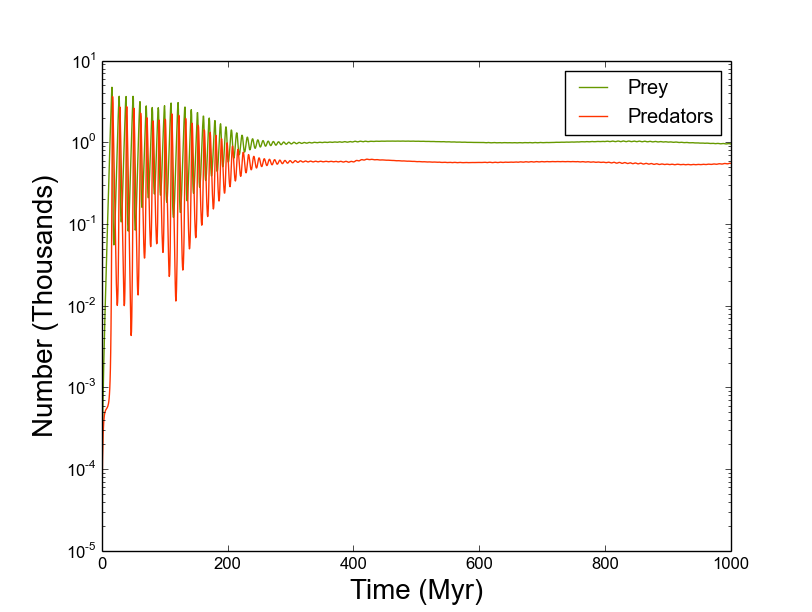}
\includegraphics[width=0.49\textwidth]{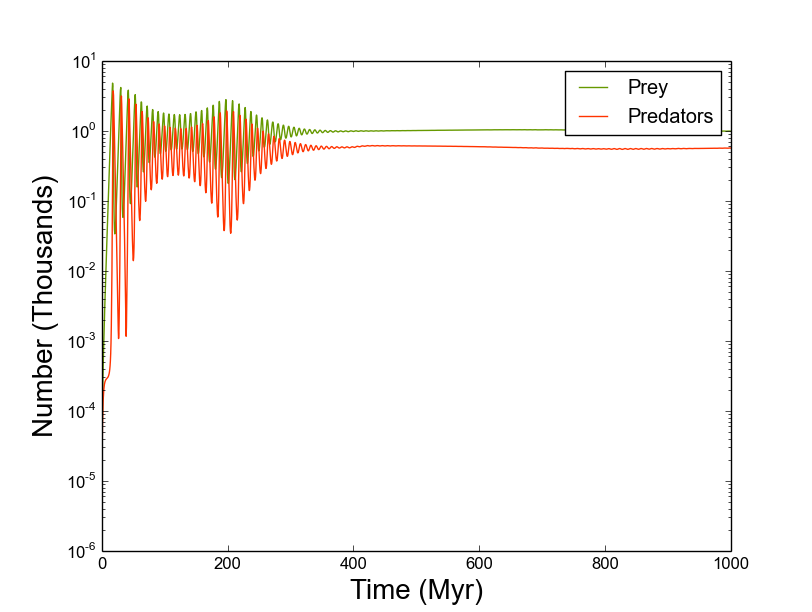}\\
\includegraphics[width=0.49\textwidth]{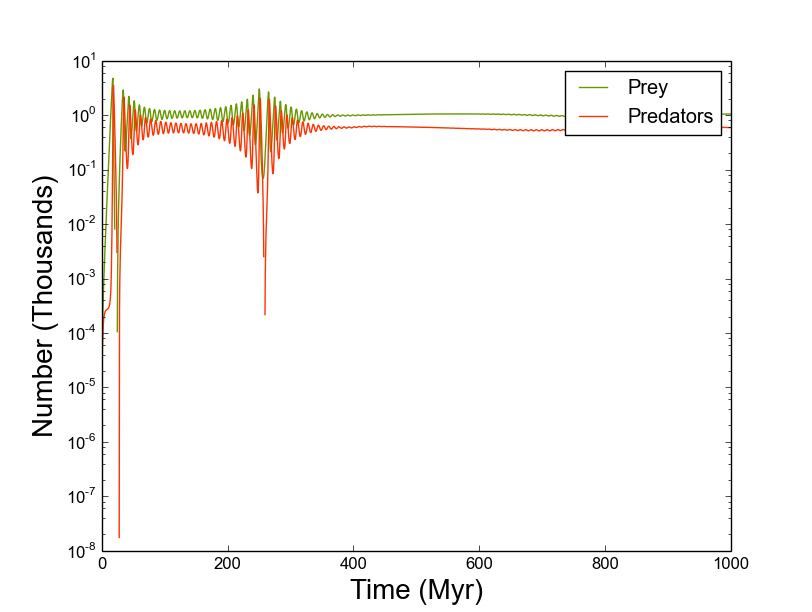}
\includegraphics[width=0.49\textwidth]{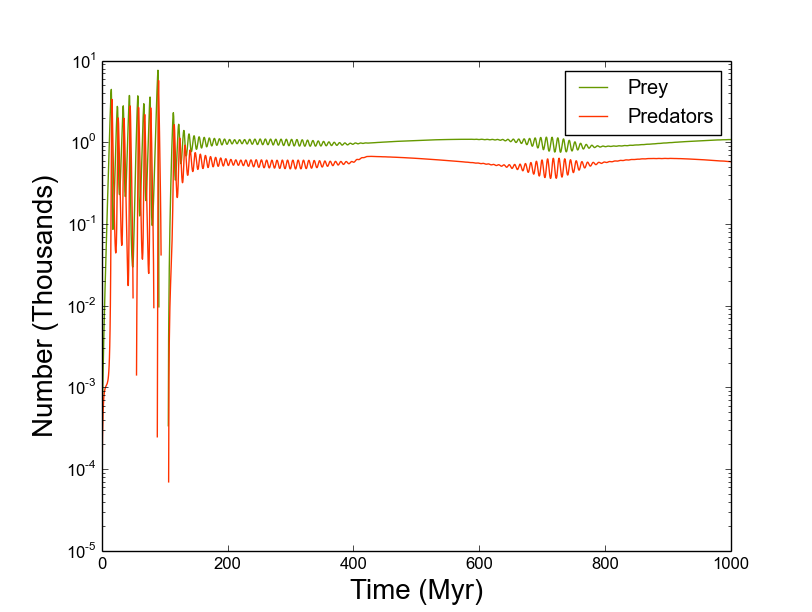}
\end{center}
\caption{The predator/prey populations of selected stars, where the outflow rate is constant and very low ($o_{R,i}=10^{-9}$).}
\label{fig:select_log_verylowoutflow}
\end{figure*}

\noindent We consider the effect of outflow rate by repeating the previous experiment, with the outflow parameter reduced to very low values ($o_{R,i}=10^{-9}$).  We find that this significantly alters both the global behaviour and the behaviour of individual systems (Figures \ref{fig:log_verylowoutflow} and \ref{fig:select_log_verylowoutflow}).

The reduced outflow rate ensures that a greater time interval is required for all stars to host predator-prey systems with sufficient quantities of each population.  As a result, the initial evolution of all systems depends strongly on their local environment - the number of directly connected neighbours, and the quantities of prey/predators arriving from each neighbour.  Once each star in the entire Galaxy is sufficiently populated, the individual systems are able to attain an equilibrium state, resulting in a global equilibrium with total predator/prey counts very similar to the previous example (where $o_{R,i}=10^{-3}$).  Note that while the behaviour of each individual system is markedly different (Figure \ref{fig:select_log_verylowoutflow}), the periodic behaviour of prey and predators in any given system remains tightly coupled.

This run demonstrates the highly time dependent nature of this solution to Fermi's Paradox.  Two quite distinct phases of evolution can be characterised - a relaxation phase which persists for the first 200-400 Myr, followed by an equilibrium phase that endures beyond 400 Myr.  Given that human SETI searches span a time interval of less than $10^{-4}$ Myr, any constraint we can place on the total number of probes in a given star system will only be a brief snapshot.  As a result, our ability to use observations of any kind to constrain any of the parameters of our model will be extremely limited indeed, even with a \emph{bona fide} detection of alien SRPs.

\subsection{Spatially Varying Prey Growth Rates}

\begin{figure}
\begin{center}
\includegraphics[width=0.7\textwidth]{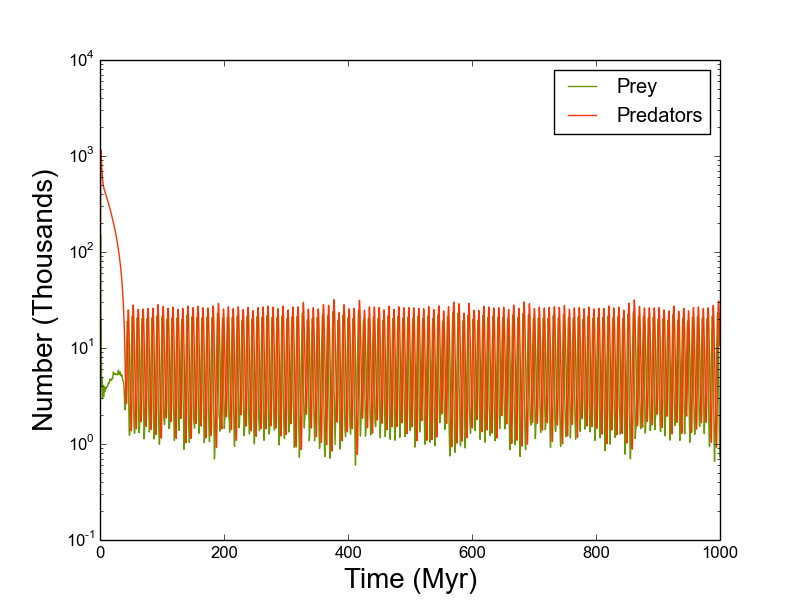}
\end{center}
\caption{As Figure \ref{fig:log_constant}, where we now allow the prey growth rate to vary uniformly amongst the 500 stars in the range $b_{R,i}=[0.5,2.0]$.}
\label{fig:log_preygrow}
\end{figure}

\noindent It is quite likely that the prey growth rate around individual stars will vary, perhaps due to the quality of resources available for self-replication.  This is likely to be a function of system chemical composition (i.e. stellar metallicity), but also the degree of element differentiation and chemical processing experienced by asteroids and minor bodies orbiting said star.  The location of the debris may also limit its usefulness to self-replicating probes - if most of the ``desirable'' raw material resides inside a deep gravitational potential well, this places energy constraints on the probe's manouevrability, further restricting its maximum replication rate.

We consider this possibility by rerunning the previous calculation (section \ref{sec:log_constant}), but now randomly sampling $b_{R,i}$ in the range $[0.5,2.0]$.  The resulting total population can be seen in Figure \ref{fig:log_preygrow}.  The steady equilibrium of the previous section has disappeared.  Periodogram analysis shows oscillatory behaviour over a range of periods, with principal period 8.77 Myr.  However, there is significant amplitude spread around this principal period, as the oscillation period of an individual system (equivalently, the velocity of the system's trajectory around its constant $H$ surface in $R-P$ surface) will depend on its individual Lotka-Volterra parameters $(b_{R,i}, d_{R,i}, b_{P,i},d_{P,i}$).  The combination of a range of oscillation periods results in a ``smearing'' of the total oscillatory period.

\subsection{Spatially Varying Prey Carrying Capacity}

\begin{figure}
\begin{center}
\includegraphics[width=0.7\textwidth]{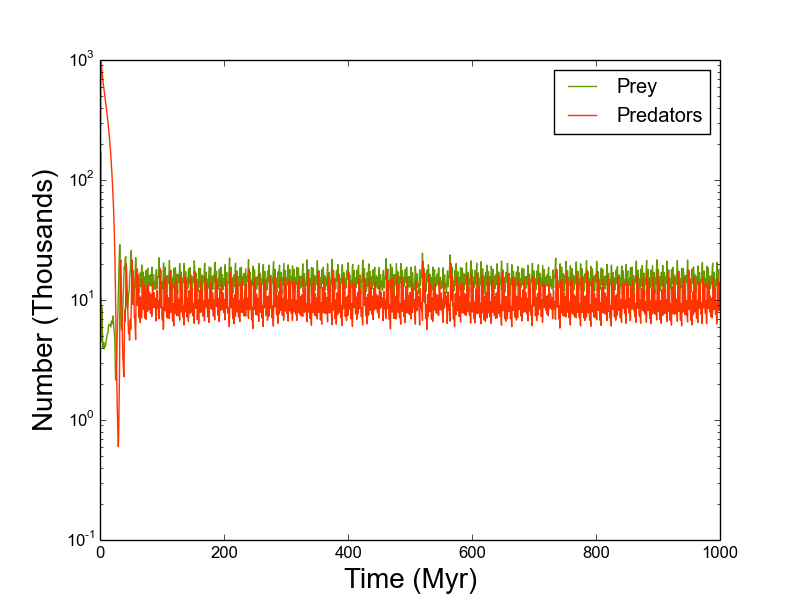}
\end{center}
\caption{As Figure \ref{fig:log_constant}, where we now allow the prey carrying capacity to vary uniformly amongst the 500 stars in the range $K_{R,i}=[5,30]$.}
\label{fig:log_carrycapacity}
\end{figure}

\begin{figure*}
\begin{center}
\includegraphics[width=0.49\textwidth]{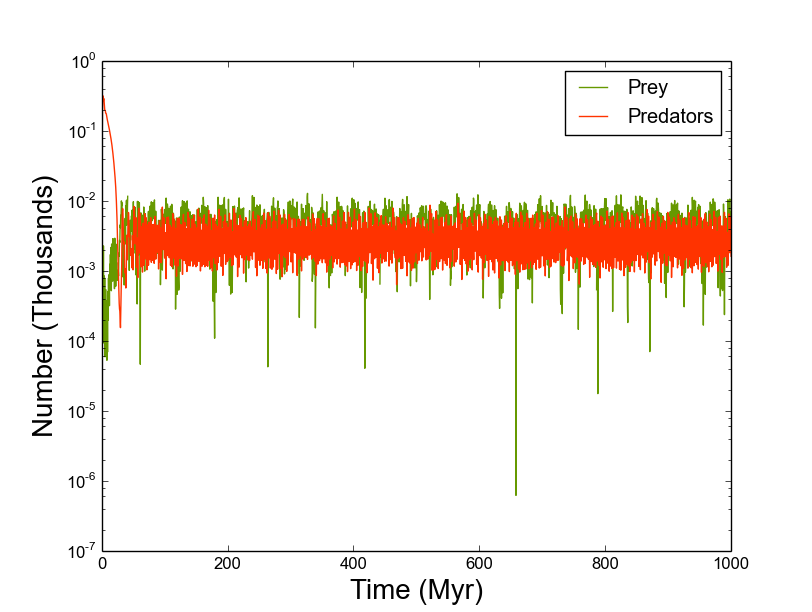}
\includegraphics[width=0.49\textwidth]{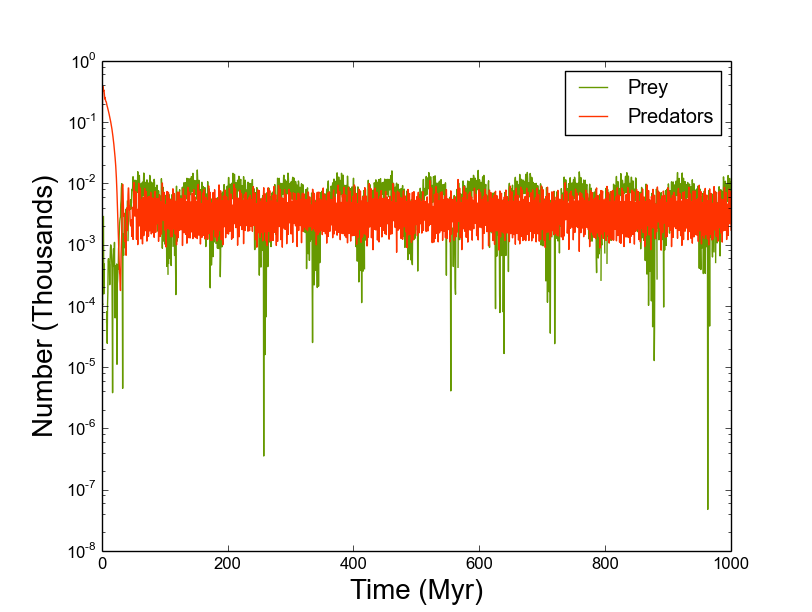}\\
\includegraphics[width=0.49\textwidth]{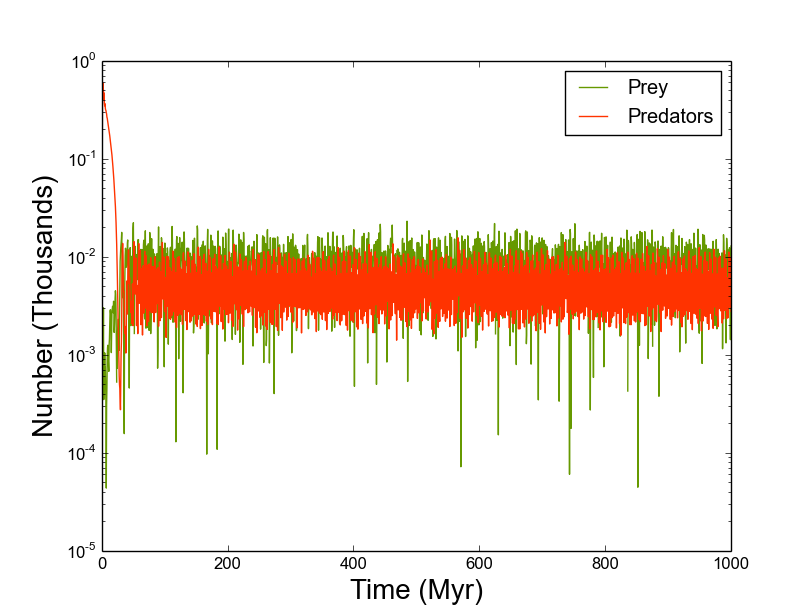}
\includegraphics[width=0.49\textwidth]{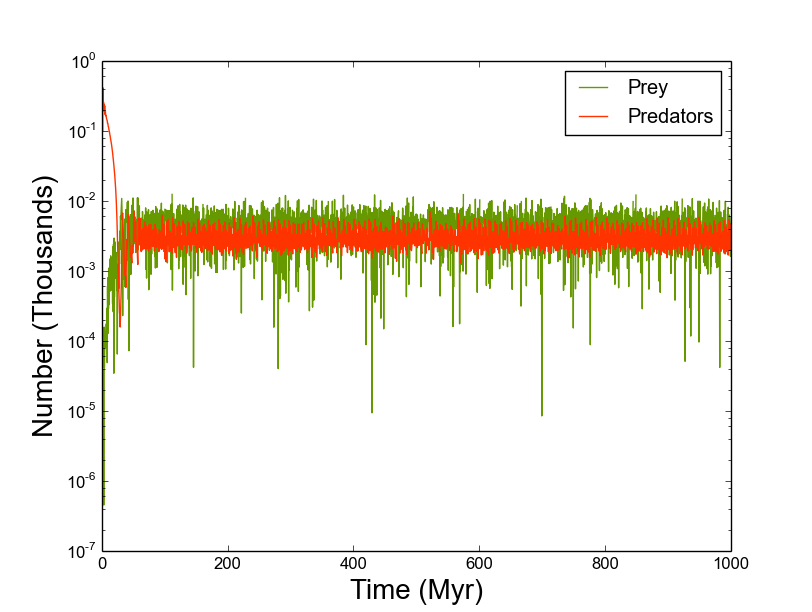}
\end{center}
\caption{The predator/prey populations of selected stars, where the prey carrying capacity varies uniformly in the range $K_{R,i}=[5,30]$.}
\label{fig:select_log_carrycapacity}
\end{figure*}

\noindent In a similar vein to the previous section, the resources provided by local asteroid belts may support varying levels of prey populations.  We might therefore expect that the prey carrying capacity will vary between individual stars.  We investigate this possibility by resetting $b_{R,i}=0.6$ and randomly sampling $K_{R,i}$ from a uniform distribution in the range $[5,30]$.  The resulting total populations again assume an oscillatory state, but with no clear principal period, and reduced variations in the value of $R$ and $P$ (Figure \ref{fig:log_carrycapacity}).  The total prey population tends to remain at larger values than the previous case, mostly because the prey growth rates can now be set at a relatively large value.

When we consider individual star systems (Figure \ref{fig:select_log_carrycapacity}), we can see that the individual predator populations are difficult to distinguish from the total predator population.  However, we can see that the prey populations show a variety of periodicities, defined not only by the local $K_{R,i}$ but also the carrying capacity of its neighbouring stars.  For example, the prey population in the top right panel of Figure \ref{fig:log_carrycapacity} exhibits a visible periodicity of around 100 Myr, whereas the bottom right panel shows no obvious evidence of periodicity.  

Notably, in contrast to every other simulation conducted so far, computing periodograms for individual systems reveals that predator and prey populations no longer share the same overall periodicity.  This decoupling is a direct consequence of the predator populations practising exponential growth, while the prey populations exercise logistic growth with varying $K$.  

\subsection{Spatially Varying Outflow Rates}

\begin{figure}
\begin{center}
\includegraphics[width=0.7\textwidth]{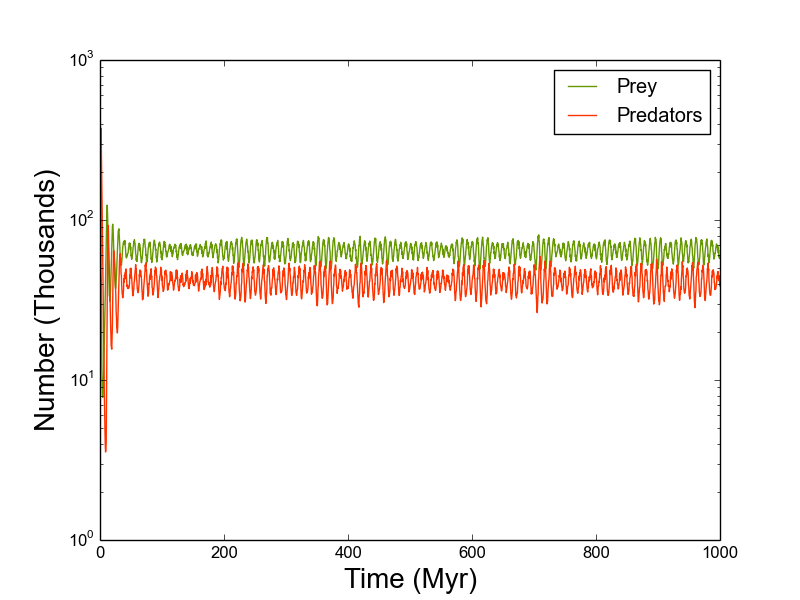}
\end{center}
\caption{As Figure \ref{fig:log_constant}, where we now allow the outflow parameter $o_{R,i}$ to vary in the range $[10^{-4},10^{-3}]$.}
\label{fig:log_varyoutflow}
\end{figure}

\begin{figure*}
\begin{center}
\includegraphics[width=0.49\textwidth]{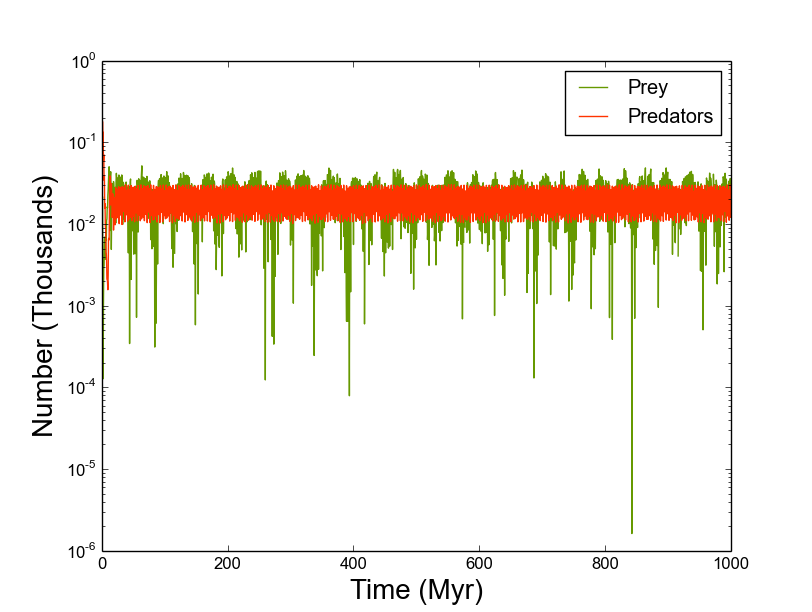}
\includegraphics[width=0.49\textwidth]{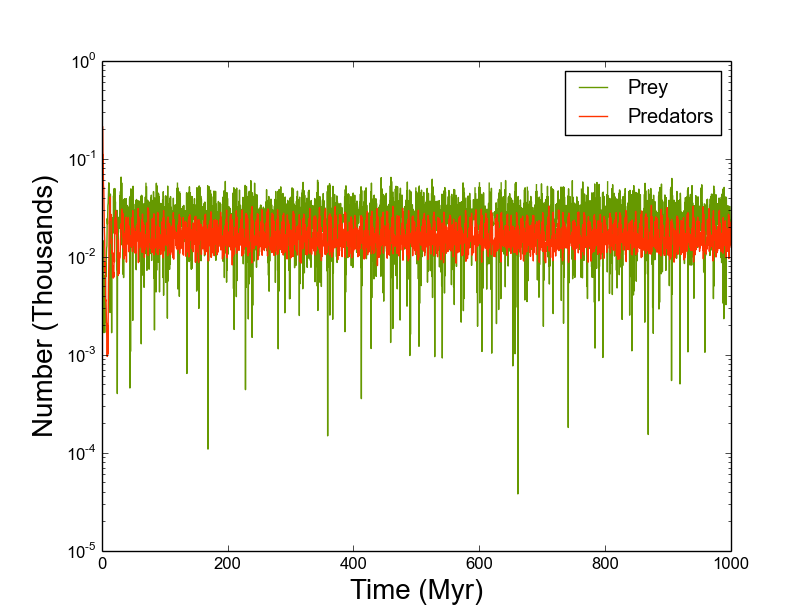}\\
\includegraphics[width=0.49\textwidth]{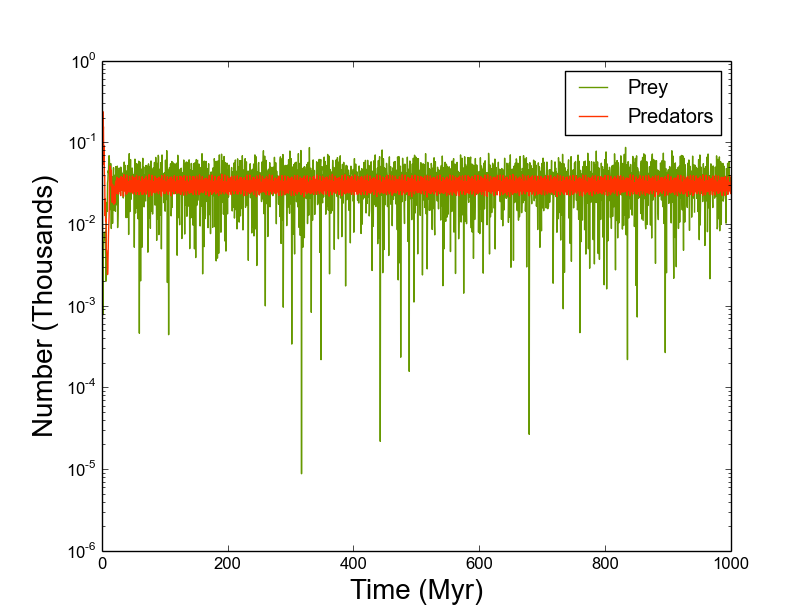}
\includegraphics[width=0.49\textwidth]{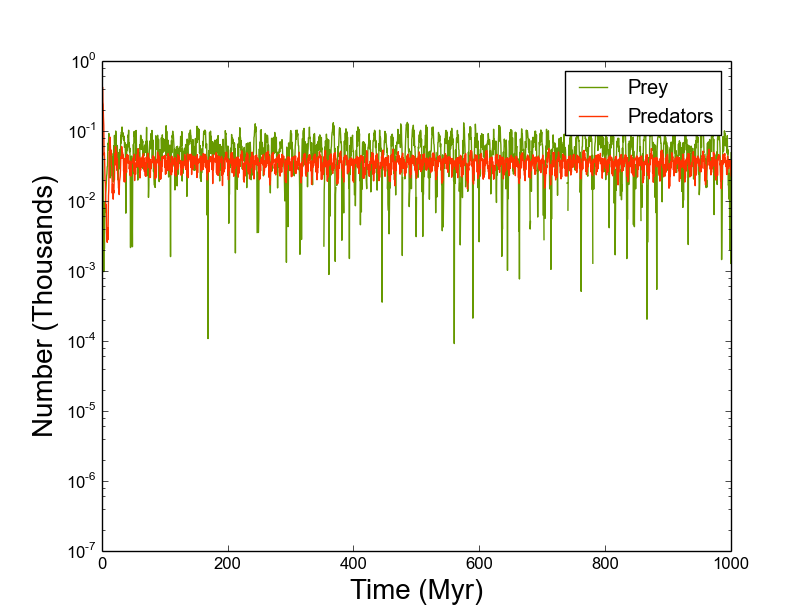}
\end{center}
\caption{The predator/prey populations of selected stars, where the outflow parameter $o_{R,i}$ varies uniformly in the range $[10^{-4},10^{-3}]$..}
\label{fig:select_log_varyoutflow}
\end{figure*}

\noindent Finally, we consider the effect of varying outflow rates between star systems by running the model with $o_{R,i}$ uniformly sampled in the range $(10^{-4},10^{-3})$ (Figures \ref{fig:log_varyoutflow} and \ref{fig:select_log_varyoutflow}).  As with the previous section, this variation forces the predator and prey populations of individual stars to oscillate on differing periods, with the oscillation period being sensitive to the local $o$ and the value of $o$ for connected neighbours.  However, Figure \ref{fig:log_varyoutflow} shows that this decoupling at local scales is not evident at global scales.  Periodogram analysis confirms that the predator and prey populations on Galaxy-wide scales oscillate with identical periodicity!

\section{Discussion}


\subsection{Comparison with other studies of Predator-Prey Dynamics}

\noindent Predator-prey dynamics has a rich history of study in mathematical biology.  The Lotka-Volterra equations have been applied to a variety of cases, in particular considering how the system is generalised to more than two species, as an attempt to model the food chain of an ecosystem (for examples see \citealt{Macarthur1967a,Smale1976,Palamara2011,Gavina2018}).

At a basic level, one can incorporate spatial effects into the vanilla Lotka-Volterra system by recasting it as a set of reaction-diffusion equations \citep{Cross1993}, where one can think of predators and prey as two reactants combining to form a product, and both entities diffuse spatially with a diffusion constant $D$, e.g.:

\begin{eqnarray}
\frac{dR(x,t)}{dt} = D_R \frac{d^2R}{dx^2} + b_{R} R - d_{R} R P \\
\frac{dP(x,t)}{dt} = D_P \frac{d^2P}{dx^2} + b_{P} P R - d_{P} P \\
\end{eqnarray}

\noindent Such systems can be stable in the absence of diffusion ($D=0$), only to become unstable when diffusion is added \citep{Turing1952}.  Of course, these models are inherently symmetric, and cannot account for spatial heterogeneities.

Most modern attempts to incorporate geography into calculations of this sort assume a lattice configuration, upon which either analytic or probabilistic solutions can be obtained.  For example, \citet{Frachebourg1996} considered the behaviour of predator-prey systems on a 1D lattice.  For a two-species system, one can model the entire evolution in terms of interfaces separating species.  Over time, species tend to organise into a mosaic of alternating domains, with the size of each domain increasing linearly with time.  If the number of species exceeds 5, these domains can become ``frozen-in'' (although this depends on whether the species chain is symmetric, i.e. can species 1 eat species 2 and species 5?).

\citet{Tome2007} construct a probabilistic cellular automaton, inspired by the Lotka-Volterra equations, on a 2D lattice.  They are able to show that self-sustained stable oscillations can be set up in the system, just as in the ``vanilla'' Lotka-Volterra case.  However, these oscillations are stable against changes in the initial conditions\footnote{see also \citealt{Nowak1992}, who derive related results on a 2D lattice using evolutionary game theory}.  A more generalised version of this result was obtained by \citet{Rozhnova2010}, who considered random networks with $k$ neighbours per node (see also \citealt{Ohtsuki2006}).

As far as the author is aware, there are no examples of a coupled Lotka-Volterra system computed on spatial graphs/networks (although see \citealt{Palamara2011} for an example on a network of \emph{species}, to resemble food webs). Lotka-Volterra systems usually model spatially continuous environments, although deliberately inserting heterogeneity into a continuous environment has been shown to remove the sustained oscillations, as we have found in our analysis \citep{McLaughlin1991, Tauber2011}.

\subsection{Limitations of the Analysis}

\noindent The flow of predators/prey between star systems depends heavily on the distance between them.  In our model, we have kept stellar positions fixed.  If we allowed the stars to move, we can expect that this will result in quasi-periodic forcing of the flow rates.  The periodicity of both predator and prey populations for a given star would be further modified, to accommodate both the star's motion and that of its neighbours.

We also assume that each star system has an unending supply of resources.  While the growth of prey is generally limited by the local carrying capacity $K$, we have not considered the possibility that replication eventually exhausts the local supply of raw materials.  One could model this rather simply as a non-constant $K$ that decays with time.  If resources could be exhausted sufficiently quickly, that might limit the number of probes overall, and provide a resolution to Fermi's Paradox.  However, if one considers the number of probes that can be produced from the available debris mass in the Solar system, the exhaustion timescale of a typical star system is likely to be much too long to be of concern to this analysis. 

That being said, we might note that the quality of raw material can vary from system to system, as a function of local metallicity.  We have attempted to model this by allowing $K$ to vary between stars.  However, our models allowed $K$ to be effectively random.  In practice, $K$ should vary according to the metallicity gradient of the Galaxy \citep[e.g.][]{Bergemann2014}.  This uniform variation may result in similar spatial variations in system periodicities.

It is also worth noting that predator probes can also scavenge other predators for resources to self-replicate.  Adding such ``omnivorous'' probes to a star system could have important consequences - reducing the predator population in this way could allow prey populations to grow to larger values.

In any case, the above additions to the analysis will not affect the final result - the predator/prey solution to the SRP formulation of Fermi's Paradox does not significantly reduce the SRP population, and is therefore not a viable solution.


\section{Conclusions}

\noindent In this paper, we have considered a proposed solution to Fermi's Paradox regarding the growth and spread of self-replicating interstellar probes.  It has been proposed that if some self-replicating probes were to ``mutate'' and begin predating other probes, this would reduce the total population of probes, ensuring that humanity would not see them.

We conduct simulations of predator-prey probe evolution using the Lotka-Volterra equations, amongst a connected network of stars in the Galactic Habitable Zone.  We find that traditional competition can result in oscillating behaviour for the predator/prey populations at a given star, as well as equilibrium solutions where both local and global populations tend to fixed values.  The nature of the system behaviour depends sensitively on the birth and death rates of each species, as well as the local carrying capacity and the flow of species between star systems.  In any case, we find that significant quantities of prey probes can persist throughout the Galaxy - admittedly less than the maximum permitted by carrying capacity, but still sufficiently large that this solution to Fermi's Paradox is weak at best, and in effect not a solution at all.

In summary, the self-replicating probe formulation of Fermi's Paradox remains, in our view, one of the strongest and most testing formulations, and an important check on our assumptions regarding the number of intelligent species in the Milky Way.

\section{Acknowledgements}

\noindent The author gratefully acknowledges support from the ECOGAL project, grant agreement 291227, funded by the European Research Council (ERC) under ERC-2011-ADG.   This research  has  made  use  of  NASA's  Astrophysics  Data  System Bibliographic  Services.  The code used in this paper is available at \url{https://github.com/dh4gan/lotka-volterra-probes}

\bibliographystyle{mnras} 
\bibliography{lotka_volterra_probes}

\begin{thebibliography}{}
\makeatletter
\relax
\def\mn@urlcharsother{\let\do\@makeother \do\$\do\&\do\#\do\^\do\_\do\%\do\~}
\def\mn@doi{\begingroup\mn@urlcharsother \@ifnextchar [ {\mn@doi@}
  {\mn@doi@[]}}
\def\mn@doi@[#1]#2{\def\@tempa{#1}\ifx\@tempa\@empty \href
  {http://dx.doi.org/#2} {doi:#2}\else \href {http://dx.doi.org/#2} {#1}\fi
  \endgroup}
\def\mn@eprint#1#2{\mn@eprint@#1:#2::\@nil}
\def\mn@eprint@arXiv#1{\href {http://arxiv.org/abs/#1} {{\tt arXiv:#1}}}
\def\mn@eprint@dblp#1{\href {http://dblp.uni-trier.de/rec/bibtex/#1.xml}
  {dblp:#1}}
\def\mn@eprint@#1:#2:#3:#4\@nil{\def\@tempa {#1}\def\@tempb {#2}\def\@tempc
  {#3}\ifx \@tempc \@empty \let \@tempc \@tempb \let \@tempb \@tempa \fi \ifx
  \@tempb \@empty \def\@tempb {arXiv}\fi \@ifundefined
  {mn@eprint@\@tempb}{\@tempb:\@tempc}{\expandafter \expandafter \csname
  mn@eprint@\@tempb\endcsname \expandafter{\@tempc}}}

\bibitem[\protect\citeauthoryear{Annis}{Annis}{1999}]{Annis}
Annis J.,  1999, J. Br. Interplanet. Soc., 52, 19

\bibitem[\protect\citeauthoryear{Balbi \& Tombesi}{Balbi \&
  Tombesi}{2017}]{Balbi2017}
Balbi A.,  Tombesi F.,  2017, Scientific Reports, 7, 16626

\bibitem[\protect\citeauthoryear{Bergemann et~al.,}{Bergemann
  et~al.}{2014}]{Bergemann2014}
Bergemann M.,  et~al., 2014, Astronomy {\&} Astrophysics, 565, A89

\bibitem[\protect\citeauthoryear{Brin}{Brin}{1983}]{BrinG.D.1983}
Brin G.~D.,  1983, QJRAS, 24, 283

\bibitem[\protect\citeauthoryear{Chyba \& Hand}{Chyba \&
  Hand}{2005}]{Chyba2005}
Chyba C.~F.,  Hand K.~P.,  2005, ARA{\&}A, 43, 31

\bibitem[\protect\citeauthoryear{{\'{C}}irkovi{\'{c}}}{{\'{C}}irkovi{\'{c}}}{2009}]{Cirkovic2009}
{\'{C}}irkovi{\'{c}} M.~M.,  2009, Serbian Astronomical Journal, 178, 1

\bibitem[\protect\citeauthoryear{Collins}{Collins}{2008}]{Collins2008}
Collins S.~G.,  2008, {All tomorrow's cultures: Anthropological engagements
  with the future}.
Berghahn Books

\bibitem[\protect\citeauthoryear{Cross \& Hohenberg}{Cross \&
  Hohenberg}{1993}]{Cross1993}
Cross M.~C.,  Hohenberg P.~C.,  1993, Reviews of Modern Physics, 65, 851

\bibitem[\protect\citeauthoryear{Denning}{Denning}{2011}]{Denning2011}
Denning K.,  2011, Acta Astronautica, 68, 381

\bibitem[\protect\citeauthoryear{Forgan}{Forgan}{2017}]{Forgan2017d}
Forgan D.~H.,  2017, International Journal of Astrobiology, 16, 349

\bibitem[\protect\citeauthoryear{Frachebourg, Krapivsky  \&
  Ben-Naim}{Frachebourg et~al.}{1996}]{Frachebourg1996}
Frachebourg L.,  Krapivsky P.~L.,   Ben-Naim E.,  1996, Physical Review E, 54,
  6186

\bibitem[\protect\citeauthoryear{Freitas}{Freitas}{1983}]{Freitas1983}
Freitas R.~A.,  1983, British Interplanetary Society, 36, 501

\bibitem[\protect\citeauthoryear{Gavina et~al.,}{Gavina
  et~al.}{2018}]{Gavina2018}
Gavina M. K.~A.,  et~al., 2018, Scientific Reports, 8, 1198

\bibitem[\protect\citeauthoryear{Gowanlock, Patton  \& McConnell}{Gowanlock
  et~al.}{2011}]{Gowanlock2011}
Gowanlock M.~G.,  Patton D.~R.,   McConnell S.~M.,  2011, Astrobiology, 11, 855

\bibitem[\protect\citeauthoryear{Haqq-Misra \& Kopparapu}{Haqq-Misra \&
  Kopparapu}{2012}]{Haqq-Misra2012}
Haqq-Misra J.,  Kopparapu R.~K.,  2012, Acta Astronautica, 72, 15

\bibitem[\protect\citeauthoryear{Lempert}{Lempert}{2014}]{Lempert2014}
Lempert W.,  2014, Visual Anthropology Review, 30, 164

\bibitem[\protect\citeauthoryear{Macarthur \& Levins}{Macarthur \&
  Levins}{1967}]{Macarthur1967a}
Macarthur R.,  Levins R.,  1967, The American Naturalist, 101, 377

\bibitem[\protect\citeauthoryear{McLaughlin \& Roughgarden}{McLaughlin \&
  Roughgarden}{1991}]{McLaughlin1991}
McLaughlin J.~F.,  Roughgarden J.,  1991, Theoretical Population Biology, 40,
  148

\bibitem[\protect\citeauthoryear{Murray}{Murray}{2004}]{Murray2004}
Murray J.~D.,  2004, {Mathematical Biology}.
 Interdisciplinary Applied Mathematics Vol. 17, Springer New York, New York, NY

\bibitem[\protect\citeauthoryear{Nicholson \& Forgan}{Nicholson \&
  Forgan}{2013}]{Nicholson2013}
Nicholson A.,  Forgan D.,  2013, International Journal of Astrobiology, 12, 337

\bibitem[\protect\citeauthoryear{Nowak \& May}{Nowak \& May}{1992}]{Nowak1992}
Nowak M.~A.,  May R.~M.,  1992, Nature, 359, 826

\bibitem[\protect\citeauthoryear{Ohtsuki \& Nowak}{Ohtsuki \&
  Nowak}{2006}]{Ohtsuki2006}
Ohtsuki H.,  Nowak M.~A.,  2006, Journal of Theoretical Biology, 243, 86

\bibitem[\protect\citeauthoryear{Ostlie \& Carroll}{Ostlie \&
  Carroll}{1996}]{Ostlie_and_Caroll}
Ostlie D.,  Carroll B.,  1996, {An Introduction to Modern Stellar
  Astrophysics}.
Pearson Education

\bibitem[\protect\citeauthoryear{Palamara, Zlatic, Scala  \&
  Caldarelli}{Palamara et~al.}{2011}]{Palamara2011}
Palamara G.~M.,  Zlatic V.,  Scala A.,   Caldarelli G.,  2011, Advances in
  Complex Systems, 14, 635

\bibitem[\protect\citeauthoryear{Papagiannis}{Papagiannis}{1978}]{Papagiannis1978}
Papagiannis M.~D.,  1978, QJRAS, 19

\bibitem[\protect\citeauthoryear{Rozhnova \& Nunes}{Rozhnova \&
  Nunes}{2010}]{Rozhnova2010}
Rozhnova G.,  Nunes A.,  2010, The European Physical Journal B, 74, 235

\bibitem[\protect\citeauthoryear{Sagan \& Newman}{Sagan \&
  Newman}{1983}]{Sagan1983}
Sagan C.,  Newman W.~I.,  1983, QJRAS, 24, 113

\bibitem[\protect\citeauthoryear{Smale}{Smale}{1976}]{Smale1976}
Smale S.,  1976, Journal of Mathematical Biology, 3, 5

\bibitem[\protect\citeauthoryear{Sotos}{Sotos}{2017}]{Sotos2017}
Sotos J.~G.,  2017, arXiv e-print 1709.01149

\bibitem[\protect\citeauthoryear{T{\"{a}}uber}{T{\"{a}}uber}{2011}]{Tauber2011}
T{\"{a}}uber U.~C.,  2011, Journal of Physics: Conference Series, 319, 012019

\bibitem[\protect\citeauthoryear{Tipler}{Tipler}{1980}]{Tipler1980}
Tipler F.~J.,  1980, QJRAS, 21, 267

\bibitem[\protect\citeauthoryear{Tom{\'{e}} \& de Carvalho}{Tom{\'{e}} \&
  de~Carvalho}{2007}]{Tome2007}
Tom{\'{e}} T.,  de Carvalho K.~C.,  2007, Journal of Physics A: Mathematical
  and Theoretical, 40, 12901

\bibitem[\protect\citeauthoryear{Turing}{Turing}{1952}]{Turing1952}
Turing A.~M.,  1952, Philosophical Transactions of the Royal Society B:
  Biological Sciences, 237, 37

\bibitem[\protect\citeauthoryear{Vukotic \& {\'{C}}irkovi{\'{c}}}{Vukotic \&
  {\'{C}}irkovi{\'{c}}}{2007}]{Vukotic_and_Cirkovic_07}
Vukotic B.,  {\'{C}}irkovi{\'{c}} M.~M.,  2007, Serbian Astronomical Journal,
  175, 45

\bibitem[\protect\citeauthoryear{Vukotic \& Cirkovic}{Vukotic \&
  Cirkovic}{2008}]{Vukotic_and_Cirkovic_08}
Vukotic B.,  Cirkovic M.~M.,  2008, Serbian Astronomical Journal, 176, 71

\bibitem[\protect\citeauthoryear{Wiley}{Wiley}{2011}]{Wiley2011}
Wiley K.~B.,  2011, arXiv e-prints 1111.6131

\makeatother
\end{thebibliography}

\end{document}